\documentclass{article}
\usepackage{spconf,amsmath,graphicx}

\usepackage{caption} 
\usepackage{tabularx}
\usepackage{float}
\usepackage{array}  
\usepackage{booktabs} 
\usepackage{multirow}
\usepackage{makecell}

\usepackage{color}

\usepackage{bbding}

\usepackage{amsfonts,amssymb} 
\usepackage{bbding}  
\usepackage{caption} 
\usepackage{tabularx} 
\usepackage{float}
\usepackage{array}  
\usepackage{booktabs} 
\usepackage{multirow}
\usepackage{makecell} 
\usepackage{color}

\usepackage{subfigure} 

\title{Multi-dimensional Edge-based Audio Event Relational Graph Representation Learning for Acoustic Scene Classification}

\name{Yuanbo Hou$^1$, Siyang Song$^2$, Chuang Yu$^3$, Yuxin Song$^4$, Wenwu Wang$^5$, Dick Botteldooren$^1$}
\address{$^1$Ghent University, Belgium.   \quad\quad
$^2$University of Cambridge, UK.\\
$^3$University of Manchester, UK.  \quad
$^4$Baidu Inc., China. \quad
$^5$University of Surrey, UK.}

\begin{document}

\captionsetup{font={small}}

\maketitle 

\begin{abstract}

\vspace{-0.1cm}
\noindent Most existing deep learning-based acoustic scene classification (ASC) approaches directly utilize representations extracted from spectrograms to identify target scenes. 
However, these approaches pay little attention to the audio events occurring in the scene despite they provide crucial semantic information.
This paper conducts the first study that investigates whether real-life acoustic scenes can be reliably recognized based only on the features that describe a limited number of audio events. 
To model the task-specific relationships between coarse-grained acoustic scenes and fine-grained audio events, we propose an \textbf{e}vent \textbf{r}elational \textbf{g}raph representation \textbf{l}earning (ERGL) framework for ASC. 
Specifically, ERGL learns a graph representation of an acoustic scene from the input audio, where the embedding of each event is treated as a node, while the relationship cues derived from each pair of event embeddings are described by a learned multi-dimensional edge feature.
Experiments on a polyphonic acoustic scene dataset show that the proposed ERGL achieves competitive performance on ASC by using only a limited number of embeddings of audio events without any data augmentations.
The validity of the proposed ERGL framework proves the feasibility of recognizing diverse acoustic scenes based on the event relational graph.
Our code is available on our homepage
{{(\textcolor{blue}{\underline{https://github.com/Yuanbo2020/ERGL}})}}.


\end{abstract}

\vspace{-0.1cm}

\vspace{-0.4cm}
\section{Introduction}
\label{sec:intro}

\vspace{-0.2cm}
Acoustic scene classification (ASC) aims to classify an audio clip from various sources in real scenarios into a predefined semantic label (e.g., park, mall, or bus) \cite{acoustic_scene}. 
ASC provides a broad description of the surrounding environment to assist intelligent agents in quickly understanding the general picture of the environment, and thus is beneficial for various applications, such as sound source recognition \cite{sound_source}, elderly well-being assistance \cite{well-being}, and audio-visual scene recognition \cite{mmsp}.

Typical deep learning-based ASC methods usually consist of three steps: first, they convert the input time-domain audio stream into a time-frequency spectrogram as its acoustic features. 
Then, the obtained acoustic features are fed to neural networks to automatically generate task-orientated representations. Finally, the classifier recognizes the acoustic scene of the input audio stream based on such high-level representations. 
For example, the paper \cite{Ren2018} utilizes a CNN-based method with mel spectrograms of input audio for ASC, where attention-based pooling layers are used to reduce the dimension of the representation.
The spatial pyramid pooling approach is used by CNN in \cite{basbug2019acoustic} to provide various resolutions for ASC. 
Except for mel spectrograms, wavelet-based deep scattering spectrum \cite{li_icmew} is introduced in ASC to exploit higher-order temporal information of acoustic features by convolutional
recurrent neural networks (CRNN) with bidirectional gated recurrent units.
Given the intrinsic relationship between acoustic scenes and audio events, some studies jointly analyze scenes and events relying on multi-task learning (MTL) \cite{Bear2019TowardsJS, tonami2021joint, komatsu2020scene}.
To further mine the implicit relational information between coarse-grained scenes and embedded fine-grained events, a relation-guided ASC \cite{RGASC} is proposed to guide the model to bidirectionally fuse scene-event relations for mutually beneficial scene and event classification.

\label{ssec:figure-f}
\begin{figure*}[t] 
	\setlength{\abovecaptionskip}{0.03cm}  
	\setlength{\belowcaptionskip}{-0.6cm}   
	\centerline{\includegraphics[width = 0.9 \textwidth]{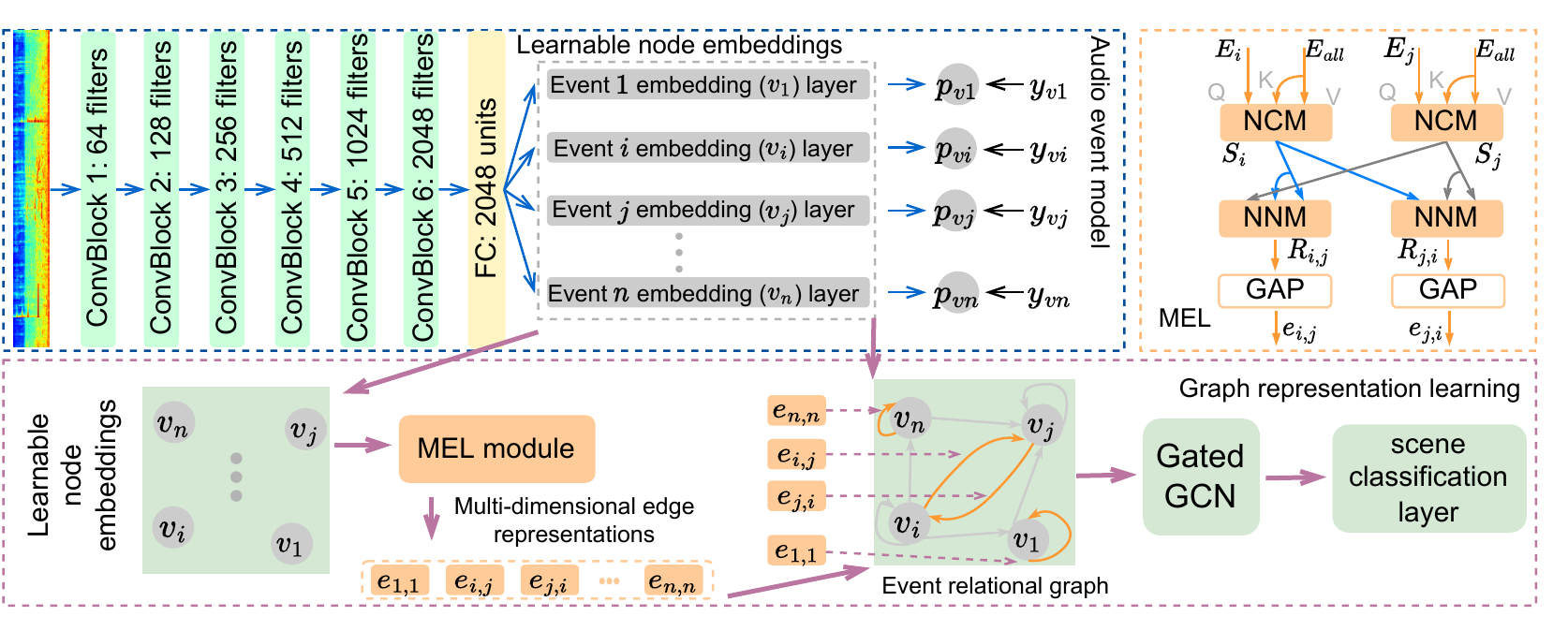}}
	\caption{The framework of scene-dependent \textbf{e}vent \textbf{r}elational \textbf{g}raph \textbf{l}earning (ERGL) for ASC.}
	\label{model}
\end{figure*}


However, most of the aforementioned approaches do not specifically consider the important semantically meaningful information in the acoustic scene (i.e., audio events). 
It is difficult to explain what types of cues in the audio stream are utilized by these approaches to recognize the acoustic scene. 
Meanwhile, it is natural for humans to recognize acoustic scenes based on the semantically meaningful audio events contained in them, where the occurring events and their relationships vary in different acoustic scenes \cite{dick}. 
This paper proposes to deep learn a pair of multi-dimensional edge-based graph to represent each audio event in an end-to-end manner, which contains not only the activation of audio events (they are treated as nodes in the graph) in the audio signal but also their task-specific relationships (represented as edges).
Then, the scene-dependent event relational graph is fed into a gated graph convolutional network (Gated GCN) to extract scene-related cues for classification.
This result shows that by relying only on several explicit audio event embeddings, the proposed ERGL can successfully build scene-dependent event relational graphs and effectively distinguish scenes.
The paper is organized as follows. Section \ref{sec:format} introduces the proposed ERGL. Section \ref{sec:experiments} describes the dataset, experimental setup, and analyzes results. Section \ref{sec:CONCLUSION} draws conclusions.

\vspace{-0.3cm}
\section{Audio \textbf{e}vent \textbf{r}elational \textbf{g}raph}
\label{sec:format}


\vspace{-0.2cm}

This paper presents a novel audio scene-event relationship modeling approach that learns a unique scene-related event relational graph from audio clips in an end-to-end manner. It first learn a set of embeddings, where each embedding ($v_i$) contains the $i$-th audio event-related information, and is treated as the $i$-th node in the event relational graph. 
Inspired by \cite{luo2022learning,song2021learning}, we further learn a pair of multi-dimensional edge features $e_{i,j}$ and $e_{j,i}$  to explicitly describe multiple task-specific relationship cues between each pair of nodes $v_i$ and $v_j$. As a result, the obtained graph explicitly describes the occurrence of a set of pre-defined audio events
and task-specific relationships among them in the given audio scene. 
Finally, the obtained event relational graph consisting of $n$ nodes and $n\times n$ edges is fed into a GatedGCN \cite{gated_GCN} for ASC.

\vspace{-0.4cm}
\subsection{Audio event node feature learning}\label{sec:node_event_label}

\vspace{-0.2cm}
To obtain node embeddings that describe audio events' occurrence in the given audio clip, we propose an audio event node feature generation model derived from the PANN \cite{kong2020panns} which shows excellent performance in recognizing audio events. 
As shown in Fig.~\ref{model}, the spectrogram of the audio clip is first fed into a set of convolutional blocks (ConvBlocks), each of which contains two convolutional layers with kernels of size 3 × 3, a batch normalization, and a ReLU activation \cite{kong2020panns}. Then, a fully-connected (FC) layer with 2048 units is used to generate a joint representation for all audio events. Different from the PANN, we further employ $n$ FC layers with 64 units to individually learn $n$ embeddings, where each embedding describes a unique pre-defined audio event.

During training, each learned event embedding $v_i$ is fed into the audio event classifier to individually predict the target event's occurrence probability $p_{vi}$, where Mean Squared Error (MSE) loss is used to measure the distance between $p_{vi}$ and the label $y_{vi}$ (i.e., 
$\mathcal{L}_\text{event} = MSE(p_{vi}, y_{vi})$). 
Specifically, to train our audio event node feature generation module, 
we use the PANN that is pre-trained on Audioset \cite{aduioset}, which contains 527 classes of audio events, to generate pseudo-labels for all events. 
It produces a $527$-dimensional soft pseudo-label $y = [y_{v1}, y_{v2}, y_{v3}, ..., y_{v{527}}]$ for each audio stream, describing the occurrence probabilities of 527 classes of pre-defined audio events. 
Since 
real-world audio scene datasets rarely have all 527 classes of events,
i.e., the number of occurred events would be much less than 527, we rank all events by accumulating their occurrence probabilities in all training data, and only use a set of top-ranked (Top $n$) audio events with the highest overall probability to describe each audio scene. As a result, each audio event graph contains $n$ nodes and $n \times n$ edges.


\vspace{-0.4cm}
\subsection{Audio event relational edge feature learning}

\vspace{-0.2cm}
Once all audio event (node) representations are produced, we propose a multi-dimensional edge learning (MEL) module to explicitly extract task-specific relationships between each pair of events, i.e., we deeply learn a pair of multi-dimensional edge features to connect each pair of nodes. Here, our hypothesis is that the co-occurrence patterns of all event pairs may include key clues for audio scene classification. For a pair of nodes, the MEL module in Fig.~\ref{model} consists of a \textbf{node-context relationship modeling (NCM)} sub-module and a \textbf{node-node relationship modeling (NNM)} sub-module. NCM first learns the task-specific relationship cues between each node (event) and the global context (scene), generating a scene-aware event feature from each node. Then, NNM further models the relationship between each pair of scene-aware event features, to generate the final multi-dimensional edge feature describing the scene-aware relationship between each pair of nodes.




\textbf{NCM.} For each nodes $v_i$, NCM conducts cross-attention \cite{Transformer} between it and the global contextual representation $v_{\text{all}}$ that concatenated all node features, where nodes $v_i$ is used as the query, and $v_{all}$ is employed as the key and value.
\begin{equation}
\setlength{\abovedisplayskip}{1pt}
\setlength{\belowdisplayskip}{1pt} 
\text{NCM}(\mathbf{Q, K} )=\Phi(\mathbf{{QW}}_q\mathbf{(KW}_k)^T/\sqrt{d_{k}})\mathbf{KW}_v 
\label{self-attention}
\end{equation} 
\begin{equation}
\setlength{\abovedisplayskip}{1pt}
\setlength{\belowdisplayskip}{1pt} 
\mathcal{S}_{i}=\text{NCM}(v_i, v_{\text{all}})
\end{equation} 
where $\Phi$ is softmax, $\mathbf{W}_q$, $\mathbf{W}_k$, $\mathbf{W}_v$ are learnable weights, and $d_k$ is a factor equal to the number of $\mathbf{K}$'s channels. As a result, the obtained representations $\mathcal{S}_{i}$ ($i = 1, 2, \cdots, n$) encode scene-aware cues for each audio event.



\textbf{NNM.} After obtaining all scene-aware event features from the NCM, the NNM module further models the relationship between each pair of them, to generate final multi-dimensional edge features for the graph. In particular, the NNM consists of a cross-attention operation and global average polling (GAP) layer, which takes a pair of the scene-aware event features $\mathcal{S}_{i}$ and $\mathcal{S}_{j}$) as the input, and a pair of multi-dimensional edge features $e_{i, j}$ and $e_{j, i}$ as output. Specifically, the NNM first conducts: 
\begin{equation}
\setlength{\abovedisplayskip}{1pt}
\setlength{\belowdisplayskip}{1pt} 
\mathcal{R}_{i, j}=\text{NNM}(\mathcal{S}_{j}, \mathcal{S}_{i}), \quad \mathcal{R}_{j, i}=\text{NNM}(\mathcal{S}_{i}, \mathcal{S}_{j})
\end{equation} 
where $\mathcal{R}_{i, j}$ encodes $\mathcal{S}_{j}$-related cues in $\mathcal{S}_{i}$, and correspondingly, $\mathcal{R}_{j, i}$ encodes $\mathcal{S}_{i}$-related cues in $\mathcal{S}_{j}$. Next, $\mathcal{R}_{i, j}$ and $\mathcal{R}_{j, i}$ are fed into a GAP layer to obtain the multi-dimensional edge feature vectors $e_{i, j}$ and $e_{j, i}$.
\begin{equation}
\setlength{\abovedisplayskip}{1pt}
\setlength{\belowdisplayskip}{1pt} 
e_{i, j} = \text{GAP}(\mathcal{R}_{i, j}), 
e_{j, i} = \text{GAP}(\mathcal{R}_{j, i})
\end{equation} 
Consequently, the learned edge features $e_{i, j}$ and $e_{j, i}$ capture multiple task-specific cues for scene classification, which relate to both event nodes $v_i$ and $v_j$, respectively.



\vspace{-0.4cm}
\subsection{ASC based on event relational graph}



Once the event relational graph that contains $n$ node embeddings $v = \{v_1, v_2, \cdots, v_n \}$ and $n\times n$ multi-dimensional directed edge representations $e = \{e_{1,1}, \cdots, e_{i,j}, \cdots, e_{n,n} \}$ are obtained, we feed it to the gated graph convolution network (GCN) \cite{gated_GCN} for audio scene classification

Since the model contains $U$ GCN layers, its output is $G^U=(v^U,e^U)$, which is a graph with the same topology as $G^0$.  
The $i$-th node represents the activation state of the $i$-th event in the scene. The latent node features in $G^U$ are concatenated as the scene representation and input to the final scene classification layer.
Cross entropy (CE) is used as the loss function in ASC between the scene prediction $p_{s}$ and the scene true label ${y}_{s}$,
$\mathcal{L}_\text{scene} = CE(p_{s}, y_{s})$. 
Hence, the total loss of the proposed model is $\mathcal{L} = \mathcal{L}_\text{event} + \mathcal{L}_\text{scene}$.


In this paper, $U$ is an important parameter in graph representation learning, and $U$ defaults to 2. The effect of $U$ on the model will be explored in the experiments discussed later.

\vspace{-0.2cm}
\section{Experiments and results}
\label{sec:experiments}

\vspace{-0.2cm}
\subsection{Dataset, Baseline, Experiments Setup, and Metric}

In this paper, TUT Urban Acoustic Scenes 2018 development dataset (UAS) \cite{DCASE2018} with 8640 10-seconds clips is used. 
UAS from real life contains 10 classes of acoustic scenes totaling 24 hours. 
UAS does not contain labels for events. 
Thus, to obtain the event labels used in Sec.~\ref{sec:node_event_label}, pre-trained PANN\protect\footnote{Model Cnn14\_16k\_mAP=0.438.pth: https://zenodo.org/record/3987831} \cite{kong2020panns} is used to annotate each audio clip with 527 classes of event pseudo-labels.
In training, following \cite{DCASE2018}, 
about 30\% of training samples are assigned to form the validation set.

A typical CNN-based approach \cite{DCASE2018} to ASC is used as the baseline. 
In addition to the baseline, this paper also presents the performance of other methods based on attention, multi-temporal features, and scene-event joint learning.

Following \cite{kong2020panns}, the log mel spectrogram with 64 bins is used as the acoustic feature, which is extracted by the Short-Time Fourier Transform with a Hamming window of size 1024 and a hop size of 320 samples. 
Dropout and normalization are used in training to prevent over-fitting of the model. 
A batch size of 64 and AdamW optimizer \cite{adamw} with a learning rate of 1e-3 are used to minimize the loss. 
Systems are trained on a GPU card (Tesla V100-SXM2-32GB) for 300 epochs.
The average accuracy (Acc) \cite{acoustic_scene} is used as the performance metric.
A higher Acc indicates a better performance of scene recognition. 
Please visit our homepage given in the Abstract for more details, source code, and some visual supplements.

\vspace{-0.2cm}
\subsection{Results and Analysis}

\vspace{-0.2cm}
\textbf{Number $n$ of target audio events.}
We first evaluate the impact of the choice of $n$, i.e.  the number of top events used in the model,  on the performance of the proposed ERGL for scene classification.
To this end, Table~\ref{tab:topn} shows the results of ERGL on ASC for different values of $n$.
The accuracy of the model does not increase monotonically as $n$ increases. 
The reason may be that as the number of events $n$ increases, the  number of nodes in the graph increases linearly, but the number of edges in the graph grows in $n^2$, which sharply increases the burden for learning the multi-dimensional edge features with the MEL module. 
The increased parameters do not provide more useful information to the model, but increase the learning burden of the model and reduce its performance.

\begin{table}[b]\footnotesize 
	\setlength{\abovecaptionskip}{0cm}   
	\setlength{\belowcaptionskip}{-0.4cm}   
	\renewcommand\tabcolsep{0.8pt} 
	\centering 
\caption{\textit{Acc} (\%) of the proposed method at different top $n$ values.}
	\begin{tabular}
	{p{0.8cm}<{\centering}|
    p{1.2cm}<{\centering}
	p{2cm}<{\centering}| 
	p{0.8cm}<{\centering}|
	p{1.2cm}<{\centering}
	p{2cm}<{\centering}} 

\hline
		{\#} & Top $n$ & $Acc$  (\%) & {\#} & Top $n$ & $Acc$  (\%)  \\ 
		
		\specialrule{0em}{0.1pt}{0.1pt}
		
        \hline
        \specialrule{0em}{0.1pt}{0.1pt}

		1 & 10 &  74.75 $\pm$ 2.61  & 6 & 150 &  75.39 $\pm$ 1.83 \\ 
  2 & 25 &  \textbf{78.08} $\pm$ 2.01  & 7 & 200 &  74.66 $\pm$ 1.40 \\ 
  3 & 50 &  76.02 $\pm$ 2.08   & 8 & 250 &  74.81 $\pm$ 1.92\\ 
  4 & 75 &  75.42 $\pm$ 1.45  & 9 & 300 &  73.97 $\pm$ 2.31 \\ 
  5 & 100 &  75.30 $\pm$ 1.57  & 10 & 400 &  72.63 $\pm$ 1.93 \\ 
		\hline
	\end{tabular}
	\label{tab:topn}
\end{table}

ERGL achieves the best result when $n$ equals 25. This means that these top 25 classes of audio events can accurately and efficiently describe the 10 different classes of scenes (`airport', `bus', `metro', `metro station', `park', `public square', `shopping mall', `street pedestrian', `street traffic', `tram') in the used dataset. These top 25 classes of audio events are, in order, [`Speech', `Vehicle', `Music', `Silence', `Animal', `Train', `Bird', `Inside, small room', `Raindrop', `Insect', `Clip-clop', `Rain', `Outside, rural or natural', `Rain on surface', `Rail transport', `White noise', `Outside, urban or manmade', `Railroad car, train wagon', `Boat, Water vehicle', `Mouse', `Horse', `Tick', `Car', `Mechanical fan', `Patter'].
So $n = 25$ will be used in the following experiments.

\textbf{Number $U$ of GCN layers.} 
Table~\ref{tab:layers} explores ERGL performance under different numbers of graph convolutional layers.
The results in Table~\ref{tab:layers} illustrate that increasing the number of layers of GCN does not lead to better results. 
The reason for this may be that the 2-layer Gated GCN already achieves a good balance between model performance and computational efficiency on the graph consisting of embeddings of 25 classes of audio events in this paper, and also that adding extra layers would make the network deeper and harder to train. Subsequent experiments will continue to use $U$ equal to 2.

\vspace{-0.2cm}
\begin{table}[H]\footnotesize 
	\setlength{\abovecaptionskip}{0cm}   
	\setlength{\belowcaptionskip}{-0.8cm}  
 \setlength{\belowcaptionskip}{-0cm}  
	\renewcommand\tabcolsep{0.8pt} 
	\centering 
\caption{\textit{Acc} (\%) of the proposed method at different $U$ layers.}
	\begin{tabular}
	{p{0.8cm}<{\centering}|
    p{1.2cm}<{\centering}
	p{2cm}<{\centering}| 
	p{0.8cm}<{\centering}|
	p{1.2cm}<{\centering}
	p{2cm}<{\centering}} 

\hline
		{\#} & $U$ & $Acc$  (\%) & {\#} & $U$ & $Acc$  (\%)  \\  
		\specialrule{0em}{0.1pt}{0.1pt} 
        \hline
        \specialrule{0em}{0.1pt}{0.1pt} 
		1 & 1 &  74.53 $\pm$ 2.12  & 5 & 5 & 75.44 $\pm$ 2.03  \\ 
  2 & 2 &  \textbf{78.08} $\pm$ 2.01  & 6 & 6 &  74.91 $\pm$ 1.88 \\ 
  3 & 3 &  76.19 $\pm$ 1.86  & 7 & 7 &  74.52 $\pm$ 1.69\\  
  4 & 4 &  76.33 $\pm$ 1.95  & 8 & 8 & 74.00 $\pm$ 1.91  \\ 
  
		\hline
	\end{tabular}
	\label{tab:layers}
\end{table}

\vspace{-0.2cm}
\textbf{Ablation study of MEL.} The expected role of MEL is to exploit multi-dimensional features to represent the edge of each pair of event nodes in scene-dependent event relational graphs.
To investigate the practical efficacy of MEL, we conduct an ablation experiment on MEL under the same model structure and training conditions.
The result for ERGL without MEL is 73.55\%$\pm$1.94\%, while that of ERGL equipped with MEL is 78.08\%$\pm$2.01\%, which illustrates that MEL that provides the model with scene-dependent multi-dimensional edge features does benefit the model.

\textbf{Comparison with prior non-ensemble ASC methods\footnote{Results are from DCASE2018 T1A website. 
The proposed ERGL only uses one end-to-end model with one type of acoustic feature and without data augmentation, whereas the top 3 methods in T1A are ensembles of multiple models with multiple features, so the top 3 results are omitted in Table \ref{tab:asc_model}.}.}
Table \ref{tab:asc_model} shows the well-performing CNN \cite{DCASE2018}\cite{Kong2018} and CRNN \cite{wang2018self} with self-attention to capture global features.
Attention is also used in pooling layers to reduce the dimensionality of CNN features \cite{Ren2018}.
Furthermore, multiple layer temporal feature (MLTF) \cite{Zhang2018-svm} is used to try to capture the dynamic temporal information of the audio signal efficiently.
And wavelet-based spectrum \cite{li_icmew} is used to exploit higher-order scene information for ASC.
PANN \cite{kong2020panns} is also estimated, and is explored in 2 modes referring to transfer learning \cite{interspeech2020hyb}.
In the fixed mode, PANN's parameters are not updated during training, and the scene is classified using prior knowledge of 527 classes of events learned from Audioset.
In the fine-tuning mode, PANN's parameters are updated using existing knowledge in capturing scene information.

The fixed-mode PANN result is comparable to Baseline, indicating that PANN with just audio event knowledge has a certain discriminative ability for scenes. 
In contrast, ERGL using just audio event embedding significantly improves scene classification accuracy even though these event embeddings are learned from pseudo labels without verification.
Overall, the proposed end-to-end EGRL without any data enhancement methods shows a competitive result compared to other methods. 
This illustrates that ERGL proposed in this paper can effectively discriminate different scenes by relying only on event embeddings, proving that a scene-dependent event relational graph built solely on the semantic information of events to identify different scenes is effective.

\begin{table}[t] \footnotesize 
\setlength{\abovecaptionskip}{0cm}   
	\setlength{\belowcaptionskip}{0cm}
	\renewcommand\tabcolsep{1pt} 
	\centering
	\caption{Comparison of non-ensemble systems on UAS dataset.}
	\begin{tabular}{
	p{3.2cm}<{\centering}|
	p{3.9cm}<{\centering}|
	p{1.0cm}<{\centering}
	} 
	    \hline 
		System & Model structure & \textsl{Acc (\%)}\\ 
		\hline
		PANN \cite{kong2020panns} (Fixed mode) & VGG-like CNN & 56.9 \\ 	
		Baseline \cite{DCASE2018} & CNN & 59.7 \\ 
		
		CNN\_from\_Surrey \cite{Kong2018} & CNN &  68.0 \\ 
		NNF\_CNNEns \cite{Nguyen2018a} & CNN and nearest neighbor filters & 69.3 \\  
		Model with Attention \cite{wang2018self} & CRNN with Self-attention & 70.8 \\ 
  ABCNN \cite{Ren2018} & Attention-Based CNN & 72.6 \\ 
		PANN (Fine-tuning mode) & VGG-like CNN & 73.8 \\ 
		MLTF \cite{Zhang2018-svm} & CNN and SVM & 75.3 \\
  Wavelet-based  spectrum \cite{li_icmew} & CRNN & 76.6 \\
  Proposed ERGL & CNN and Graph Learning & \textbf{78.1} \\ 
	\hline
	\end{tabular}
	\label{tab:asc_model}
\end{table}

\textbf{Comparison with scene-event joint analysis methods.}
The ERGL proposed in this paper infers target scenes based on the implicit event relational graph in the scene, which is a sene-event joint analysis method. Table \ref{tab:asc_joint} compares ERGL with  other scene-event joint analysis methods.
In Table \ref{tab:asc_joint}, \#1 based on the same latent space for scene and event classification performed the worst. The reason may be that real-life coarse-grained scenes and fine-grained events differ at the semantic level and the feature space.
Based on MTL, \cite{tonami2021joint} attempts to use shared base features and separated task-dependent high-level features to identify scenes and events.
\cite{komatsu2020scene} is based on a conditional loss from one-way scene-event to analyze scenes and events jointly.
RGASC \cite{RGASC} exploits the scene-event relationship to guide the two-tower model to bidirectionally fuse scene-event information to achieve mutually beneficial scene-event classification.
The proposed ERGL relies only on the representation of events to achieve one-way event-to-scene inference, and deduces target scenes based on the corresponding implicit event relational graph.
It is worth noting that ERGL, which only needs 25 classes of audio event information, outperforms RGASC using 527 classes of audio event information, which shows the effectiveness of ERGL.
Overall, 
the proposed ERGL achieves competitive and promising results, demonstrating the feasibility of inferring acoustic scenes based on the event relational graph.

\vspace{-0.2cm}
\begin{table}[H]\footnotesize
\setlength{\abovecaptionskip}{0cm}   
\setlength{\belowcaptionskip}{0.cm}
\renewcommand\tabcolsep{1pt} 
	\centering
	\caption{ASC results of scene-event joint analysis methods.}
	\begin{tabular}{
	p{0.7cm}<{\centering}|
	p{6.cm}<{\centering}|
	p{1.4cm}<{\centering}
	}
	    \hline
		\# & Method &  \textsl{Acc} (\%)\\
		\hline
		1 & Scene and event jointly classification \cite{Bear2019TowardsJS} &   52.35 \\ 
		
		2 & MTL-based event and scene analysis   \cite{tonami2021joint} &    61.69 \\
		
		3 & Conditional scene and event recognition \cite{komatsu2020scene} &   66.39 \\ 
		
		4 & Relation-guided ASC (RGASC) \cite{RGASC} &  77.35 \\

  5 & The proposed ERGL &  \textbf{78.08} \\
	
	\hline
	\end{tabular}
	\label{tab:asc_joint}
\end{table}

\vspace{-0.6cm}
\section{CONCLUSION}
\label{sec:CONCLUSION}

\vspace{-0.2cm} 
This paper explores the feasibility of building scene-dependent event relational graphs to recognize various acoustic scenes. 
The proposed ERGL represents each acoustic scene by learning a set of audio event embeddings as graph node features and specifically producing a pair of scene-aware multi-dimensional edge features to describe the event relationships. 
Experimental results show that ERGL achieves competitive ASC performance on a real-life dataset.
Future work will analyze and visualize the bidirectional multi-dimensional edges learned by ERGL to clarify scene-event relationships and build a scene-event dynamic relationship model.


\vfill\pagebreak

\label{sec:refs}

\bibliographystyle{IEEEbib}
\bibliography{Template}

\end{document}